\documentclass[aps,pra,onecolumn,groupedaddress,showpacs]{revtex4}

\usepackage{hyperref}
\usepackage{graphics}
\usepackage{amssymb,amsmath}

\draft
\usepackage{hyperref}
\usepackage{graphicx}

\newcommand{\om}{\omega}

\newcommand{\ve}{\varepsilon}
\newcommand{\pa}{\partial}

\begin{document}

\title{Beer's law in semiconductor quantum dots}
\author{G. T. Adamashvili}
\altaffiliation[Permanent address: ]{Georgian Technical University, Kostava str.
77, Tbilisi, 0179, Georgia. }
\affiliation{Max-Planck-Institut f\"ur Physik Komplexer Systeme N\"othnitzer Str. 38 D-01187 Dresden, Germany\\
email:gadama@parliament.ge}

\author {A. A. Maradudin }
\affiliation{Department of Physics and Astronomy\\ and Institute for Surface and Interface Science\\
 University of California, Irvine, CA, 92697 USA.\\
email: aamaradu@uci.edu}

\begin{abstract}
The propagation of a coherent optical linear wave in an ensemble of semiconductor quantum dots is considered. It is shown that a distribution of transition dipole moments of the quantum dots  changes significantly the polarization and Beer's absorption length of the ensemble of quantum dots.
Explicit analytical expressions for these quantities are presented.
\end{abstract}


\maketitle

\section{Introduction}

Semiconductor quantum dots (SQD), also referred to as zero-dimensional systems, are nanostructures that allow confinement of  charge carriers in all three spatial directions, which results in atomic-like discrete energy spectra and  strongly enhanced carrier lifetimes. Such features make quantum dots similar to atoms in many respects (artificial atoms)\cite{1}. Observation of optical coherence effects in ensembles of quantum dots is usually spoiled by the inhomogeneous line broadening due to dot-size fluctuations, with a full width at half maximum of typically more than several tens of meV. Real quantum dots often have a base length in the range 50-400~{\AA}. Beside  the frequency, the quantum dot-size fluctuations influence also the transition dipole moments of SQD. Borri {\it  et al.}\cite{2} have reported measurements of optical Rabi oscillations in the excitonic ground-state transition  of an inhomogeneously broadened InGaAs quantum dot ensemble. They found that a distribution with a 20 percent standard deviation of transition dipole moments results in a strong damping of the oscillations versus pulse area.
In the experiments reported in \cite{2}  it was also  found that  the period of the Rabi oscillations is changed. These results show quantitatively how uniformity in dot size and shape is important for any application based on a coherent light-quantum dot ensemble interaction.

In theoretical investigations of optical coherence effects in ensembles of SQD it is usually assumed that transition dipole moments of the quantum dots are independent of the size of the dots [3-8].
It is obvious that for an adequate description of  coherent optical phenomena it is necessary to take into account a distribution of transition dipole moments in the ensemble of quantum dots.

The purpose of the present work is to  investigate theoretically the propagation and absorption of an optical linear  wave  when  a distribution of transition dipole moments in the ensemble of SQD is taken into account.

\vskip+0.5cm

\section{The optical Bloch equations in semiconductor quantum dots}

We consider the propagation of a circularly polarized optical plane wave pulse in an ensemble of SQD with a  pulse  width $T<<T_{1,2}$, frequency $\omega >>T^{-1}$, wave vector $\vec{k}$, and the electric field $\vec{E}$, which gives rise to the excitonic ground-state transitions (labeled 0-X).

We use the method of slowly varying envelopes, and define a real envelope
$\hat{E}$  by
$$
\vec{E}(z,t)=(\sqrt{2})\hat{E}(z,t)[\vec{x} \cos(\om t- kz)+\vec{y}\sin(\om t- kz)],
$$
which defines an optical pulse propagating along the positive $z$ axis, and $\vec{x}$ and $\vec{y}$ are unit vectors in the directions of the $x$ and $y$ axes. We shall find it useful to define
the complex polarization vectors
$$
\vec{e}_{+}=\frac{1}{\sqrt{2}}(\vec{x}+i\vec{y}),\;\;\;\;\;\;\;\;\;\;\;\;\;\;
\vec{e}_{-}=\frac{1}{\sqrt{2}}(\vec{x}-i\vec{y}),
$$
in terms of which  the electric field has the form
\begin{equation}
\vec{E}(z,t)=\hat{E}(z,t)(\vec{e}_{-}e^{i(\om t- kz)}+\vec{e}_{+}e^{-i(\om t- kz)} ),
\end{equation}
where  $\hat{E}(z,t)$ is the slowly varying complex envelope of the electric field and $e^{\pm i(kz -\omega t)}$ contains the rapidly varying  phase of the carrier wave. We will assume that
\begin{equation}
|\frac{\partial \hat{E}}{\partial t}|<<\omega
|\hat{E}|,\;\;\;|\frac{\partial \hat{E}}{\partial z }|<<k|\hat{E}|.
\end{equation}

For the description of SQD in a circularly polarized optical pulse
we use the two-level model (0-X transitions) that can be described by the states $|1>$ and $|2>$
with energies $E_{1}=0$ and $E_{2}=\hbar \omega_{0}$, respectively, where $|1>$ is the ground state. The Hamiltonian and wave function of this system are:
$$
H=H_{0}+\hat V,
$$
$$
|\Psi >=\sum_{n=1,2} c_{n}(t) exp(-\frac{i}{\hbar} E_{n} t )|n>,
$$
where $H_{0}=\hbar\omega_{0}|2><2|$ is the Hamiltonian of the two-level SQD with the frequency of excitation $ \omega_{21}=\omega_{0}$,
$$
\hat V=-\hat{\mu} \vec{E}=-\mu (\hat{\sigma}_{-} E_{+}+\hat{\sigma}_{+}E_{-} )
$$
is the light-SQD interaction Hamiltonian, $\hbar$ is Planck's constant,
$$
 \hat{\mu}=\vec{\mu}_{12}  \hat{\sigma}_{+}+ \vec{\mu}^{*}_{12}\hat{\sigma}_{-}
$$
is the SQD's dipole moment operator,
$$
\vec{\mu}_{12}=\frac{\mu}{\sqrt{2}}(\vec{x} - i \vec{y})
$$
is the electric dipole matrix element for the corresponding transition,
$$
\mu=|\vec{\mu}_{12}|,\;\;\;\;\;\;\;\;\vec{\mu}_{21}=\vec{\mu}^{*}_{12},
$$
$$
 {E}_{\pm} =\hat{E}e^{\pm i(\om t- kz)},
$$
$$
\hat{{\sigma}}_{\pm}=\frac{1}{2}(\hat{\sigma}_{x}\pm i\hat{\sigma}_{y}),
$$
and the $\{\hat{{\sigma}}_{i}\}$ are the Pauli matrices, which satisfy $[\hat{{\sigma}}_{x},\hat{{\sigma}}_{y}]=2i\hat{{\sigma}}_{z}$, and commutation relations resulting from cyclic permutations of the subscripts.

The probability amplitudes $c_{1}$ and $c_{2}$ are determined by the Schr\"odinger equations
\begin{equation}
i\hbar \frac{\partial c_{1}(t)}{\partial t} = -\mu \;{E}_{+} c_{2}(t)e^{-i\om_{0}t}
$$$$
i\hbar \frac{\partial c_{2}(t)}{\partial t} =-\mu\; {E}_{-} c_{1}(t)e^{i\om_{0}t}.
\end{equation}

The average values of the Pauli operators $\hat{\sigma}_{i}$ for the state $|\Psi >=c_{1}\;|1> + \; c_{2}\;|2>$, are $s_{i} = <\hat{\sigma}_{i}>\;\;$ $ = <\Psi|\hat{\sigma}_{i}|\Psi> $, (where $i=1,2,3)$ and have the form \cite{9}:
$$
s_{x}=c^{*}_{1}(t)c_{2}(t)e^{-i\om_{0}t}+c_{1}(t)
c^{*}_{2}(t)e^{i\om_{0}t},
$$$$
s_{y}= ic^{*}_{1}(t)c_{2}(t)e^{-i\om_{0}t}
-ic_{1}(t)c^{*}_{2}(t)e^{i\om_{0}t},
$$$$
s_{z}=c^{*}_{2}(t) c_{2}(t)-c^{*}_{1}(t)c_{1}(t).
$$

To introduce explicitly the appropriate rotation matrix for the functions  $s_{i}$
in the rotating frame with components $u, v, w$, from the Schr\"odinger equations (3)
we obtain the Bloch equations:
\begin{equation}
\dot{u}=-\Delta v,
$$$$
\dot{v}=i\Delta u + \kappa w \hat{E},
$$$$
\dot{w}=- \kappa v \hat{E},
\end{equation}
where
$$
\kappa=\frac{2 \mu}{\hbar},\;\;\;\;\;\;\;\;\;\;\;\;\;\;\;\;\Delta=\omega_{0}-\omega,
$$
while $u$ and $-v$ are the components, in units of the
transition moment $\mu $, of the SDQ's dipole moment in-phase and in-quadrature with the field $\vec{E}$. In other words, $v$ is the absorptive component of the SQD dipole moment, while $u$ is the dispersive component,  and $\frac{1}{2}\hbar \om_{0} s_{z}$ is the expectation of the atom's unperturbed energy.

In addition to (4), we need a description of the pulse propagation in the medium. The wave equation for the electric field $\vec{E}(z,t)$
of the optical pulse in the medium is given by
\begin{equation}
\frac{\pa^{2} \vec{ E}}{\pa {z}^2} -\frac{\ve}{c^2} \frac{\pa^{2} \vec
{E}}{\pa t^2}=\frac{4\pi}{c^2}\frac{\pa^{2}\vec {P} }{\pa t^2},
\end{equation}
where $c$ is the speed of  light  in vacuum, $\ve=\eta^{2}$ is the dielectric permittivity, $\eta$ is the refractive index of the medium, and $\vec{P}$ is the polarization of the ensemble of  SQD.

For the determination of the polarization of the ensemble of  SQD we have to take into account that the dipole moment of the SQD depends on the size of the quantum dot. The polarization of the ensemble of  SQD is equal to
$$
\vec{P}=\sum_{i=1}^{n_0}<\vec{\mu}_{i}>,
$$
where $n_{0}$ is the uniform dot density, and $<\vec{\mu}_{i}>$  is the expectation value of the i-th
dipole moment.

We can rewrite this expression in another form:
\begin{equation}
\vec{P}=\sum_{j=1}\;n_{j}\;<\vec{\mu}_{j}>,
\end{equation}
where $n_{j}$ is the number of dipoles with matrix elements $\mu_{j}$ in the interval $\Delta \mu_{j}$.

We introduce the function  $\tilde{h}(\mu_{i})$ such that the quantity $\tilde{h}(\mu_{i})\Delta \mu_{i} $ is the the fraction of dipoles with dipole matrix element $\mu_{i}$ within $\Delta \mu_{i} $.
Obviously its normalization is
$$
\int_{0}^{\infty}\tilde{h}(\mu)d \mu =1,
$$
or in another form
$$
\int_{\mu_{min}}^{\infty}\tilde{h}(\mu-\mu_{min})d \mu =1,
$$
where $\mu_{min}$ is the minimum value of the dipole matrix element in the ensemble of SQD.
To take into account that when $\mu<\mu_{min}$  the  number of
dipoles is equal to zero, $\tilde{h}(\mu<\mu_{min})=0$, we can rewrite  the last equation in the form
$$
\int_{-\infty}^{\infty}\tilde{h}(\mu-\mu_{min})d \mu =1.
$$

Because the quantity ${\mu}_{0}-\mu_{min}$ is constant, we can introduce the function
$$
{h}(\mu-{\mu}_{0})=\tilde{h}(\mu-\mu_{min}),
$$
and for this function  the normalization condition has the form
$$
\int_{-\infty}^{\infty}{h}(\mu-{\mu}_{0})d \mu =1,
$$
where
$$
\vec{\mu}_{0}=\frac{\sum_{i=1}^{n_{0}} \vec{\mu}_{i}}{n_{0}}
$$
is the main dot dipole matrix element.

By using the function $h(\mu-{\mu}_{0})$, we obtain that the quantity  ${n_{j}}\approx {n_{0}}\;{h}(\mu_{j}-\mu_{0})\;\Delta \mu_{j}$. Substituting the last expression into  Eq. (6) we obtain the polarization of the ensemble of SQD   in the  form
$$
\vec{P}\approx \;{n_{0}}\;\sum_{j=1}^{\tilde{n}_0}\;{h}(\mu_{j}-\mu_{0})\;\Delta \mu_{j}\;<\vec{\mu_{j}}>.
$$
After transition to a limit we obtain the exact value of the polarization, namely
\begin{equation}
\vec{P}= {n_{0}}\;\int_{-\infty}^{\infty}\;{h}(\mu-\mu_{0})\;\;<\vec{\mu}>\;d\mu.
\end{equation}

Expressing the value of  $<\mu>$ in Eq. (7) in terms of the slowly varying amplitudes $u$ and $v$, we obtain
$$
\vec{P}= \frac{{n_{0}}}{\sqrt{2}}\;\int_{-\infty}^{\infty}\;\mu\;{h}(\mu-\mu_{0})\;\{u \;[\vec{x} \cos(\om t - k z)\nonumber\\
+\vec{y}\sin(\om t-kz)]+v \;[-\vec{x}\sin(\om t - k z)+\vec{y}\cos(\om t - k z)]\}\;d\mu.
$$

The  fluctuations in the size of the SQD also lead to the inhomogeneous broadening of the spectral line.
Upon taking into account the inhomogeneous broadening of the spectral line, the polarization of the ensemble of SQD is equal to
\begin{eqnarray}
\vec{P}(z,t)= \frac{{n_{0}}}{\sqrt{2}}\;\int_{-\infty}^{\infty}\;\int_{-\infty}^{\infty}\;\mu\;{h}
(\mu-\mu_{0})\;g(\Delta)\{u(z,t;\Delta,\mu) [\vec{x} \cos(\om t - k z)+\vec{y}\sin(\om t-kz)]\nonumber\\
+v(z,t;\Delta,\mu)[-\vec{x}\sin(\om t - k z)+\vec{y}\cos(\om t - k z)]\}\;d\mu\;d\Delta,
\end{eqnarray}
where $g(\Delta)$ is the inhomogeneous broadening function.
This equation generalizes the polarization of the ensemble of SQD to the case of a distribution of transition dipole moments. In the special case, when all dipole moments are the same, the distribution function is $h(\mu-\mu_{0})=\delta(\mu-\mu_{0})$, and we obtain the  polarization in the usual form \cite{9}:
\begin{eqnarray}
\vec{P_{h}}(z,t)= \frac{{n_{0}}\mu_{0}}{\sqrt{2}}\;\int_{-\infty}^{\infty}\;g(\Delta)\{u(z,t;\Delta) [\vec{x} \cos(\om t - k z)+\vec{y}\sin(\om t-kz)]\nonumber\\
+v(z,t;\Delta)[-\vec{x}\sin(\om t - k z)+\vec{y}\cos(\om t - k z)]\}\;d\Delta.
\end{eqnarray}

By substituting equations (1) and (8) into the wave equation (5), and  using the assumption of slowly varying amplitudes (2), we obtain the in-phase and in-quadrature wave equations in the following forms:
\begin{equation}
(k^{2}   -\frac{\ve}{c^2}  \om^{2} )\hat{E}= \frac{2\pi n_{0} {\om}^{2}}{c^2 }\;\int_{-\infty}^{\infty}\;\int_{-\infty}^{\infty}\;
\mu\;{h}(\mu-\mu_{0})\;g(\Delta)\; u  \;d\mu\;d\Delta,
\end{equation}
\begin{equation}
 \frac{\pa \hat{E}}{\pa z}   +\frac{ \om \ve}{  kc^2}  \frac{\pa \hat{E}}{\pa t}  = \frac{\pi n_{0} {\om}^{2}}{c^2  k}\;\int_{-\infty}^{\infty}\;\int_{-\infty}^{\infty}\;
\mu\;{h}(\mu-\mu_{0})\;g(\Delta)\; v  \;d\mu\;d\Delta.
\end{equation}

Equations (10) and (11) are general equations for the slowly varying amplitudes
by means of which we can consider a quite wide class of coherent optical phenomena
(for instance: Rabi oscillations,  photon echo, self-induced transparency, and others)
in an ensemble of SQD in the presence of a distribution of transition dipole moments in the ensemble.
Equations  (10) and (11)  generalize the  Maxwell equations that have been considered up to now for atomic systems and for SQD when  fluctuations of the dipole moments are neglected \cite{1}.
We have to note that an analytic solution of these equations are significantly  difficult than the solution of the Maxwell equations in the special case, when all dipole moments are the same,  because in Eqs.(10) and (11) the quantities $v$ and $u$ are the functions also of the variable $\mu$.
Nevertheless analytic solution of the equations (10) and (11) are possible in some special cases, for instance, for Beer's law in SQD.

\vskip+0.5cm

\section{Beer's law in semiconductor quantum dots}

From Eq. (10) we can obtain the  dispersion relation for the wave and from Eq. (11) we can determine the explicit form of the envelope  of the electric field strength $\hat{E}$.

Usually $g(\Delta)$  is a very broad and smooth function, so that the integral $\int_{-\infty}^{\infty}g(\Delta) u d\Delta$ in Eq. (10) is very small and  is usually neglected \cite{9}. Consequently, the dispersion law has the form
\begin{equation}
k^{2}=\frac{\ve}{c^2}  \om^{2}.
\end{equation}

After integration of the in-quadrature wave equation (11) from $t=-\infty$ up to the time
$\bar{t}$ that occurs after the pulse has passed the point of observation $z$ we obtain
\begin{equation}
 \frac{\pa \vartheta(\bar{t},z)}{\pa z}  = \frac{2 \pi^{2} n_{0} {\om}^{2} g(0)}{c^2  k} \int_{-\infty}^{\infty}
\mu {h}(\mu-\mu_{0}) v(\mu,t_{0},z,0) d\mu,
\end{equation}
where
\begin{equation}
\vartheta(t,z)=\int_{-\infty}^{t}\;\hat{E}(t',z)dt'
\end{equation}
is  proportional to the area of the pulse, $\theta (t,z)=\frac{2\mu }{\hbar}\vartheta(t,z)$, the time $t_0$ marks the end of the pulse, and at $t_0$  and for all later times the
electric field envelope $\hat{E}$ is zero.

For an analytic solution of Eq. (13) we have to determine the explicit form of the
quantity $v$ as a function of the electric field and dipole moment. From the Bloch equations (4) for the absorptive part of the on-resonance dipole amplitude, we find that
\begin{equation}
 v(\mu,t_{0},z,0)=- \sin \frac{2 \mu}{\hbar} \vartheta(t_{0} ,z).
\end{equation}

We consider  Beer's law of absorption for the electric field of the pulse in a SQD.
For this purpose  we consider the limit of weak electric fields, in which the "area" $\theta(t_{0} ,z)$ is small, i.e. $|\theta(t_{0} ,z)|<<1$, for all values of $\mu$. Under this condition, and  taking into account that for all $t>t_0$, $\vartheta(\bar{t},z)=\vartheta(t_{0},z)$, from  Eq.(15) we obtain
\begin{equation}
 v(\mu,t_{0},z,0)\approx- \frac{2 \mu}{\hbar} \vartheta(t_{0} ,z).
\end{equation}

On substituting Eq. (16) into Eq. (13), we obtain the relation

\begin{equation}
 \frac{\pa \vartheta(\bar{t},z)}{\pa z}  =- \frac{\alpha}{2}\; \vartheta(\bar{t} ,z),
\end{equation}
where
\begin{equation}
 \alpha=\frac{4\;\pi^{2} n_{0} {\om}^{2}g(0)}{c^2  \hbar \; k}\;\;\int_{-\infty}^{\infty}\;
{\mu}^{2}\;{h}(\mu-\mu_{0})\;d\mu
\end{equation}
is the reciprocal Beer's absorption length, in which   fluctuations of the dipole moments of the quantum dots have been taken into account.

\vskip+0.5cm

\section{Conclusion}

We  note that  Beer's law, Eq.(17), shows that the electric field of the wave decays exponentially with increasing penetration into the medium. For  pulses with an electric field  of  arbitrary shape,  Beer's law is usually written  for  the dimensionless   pulse area $\frac{2 \mu_{c}}{\hbar} \vartheta $, where  the dipole moment $\mu_{c}$ is the same for all atoms or SQD.

For SQD with a distribution of their dipole moments,  it is impossible to write  the area of a pulse in the usual dimensionless form, and we have to use Beer's law in the form of  Eq. (17), for the quantity $\vartheta$ defined by Eq. (14).

In the limiting case when the fluctuation of the dipole moments is neglected, the distribution function
${h}(\mu-\mu_{0})$  transforms  to the Dirac $\delta$-function, and the reciprocal Beer's absorption length reduces to the usual form which,  in atomic systems with identical dipole moments in solids or in gases, where $\eta=1$ \cite{9}, is
\begin{equation}
 \alpha_{h}=\frac{4\;\pi^{2} n_{0}\om  g(0) }{c  \hbar \; \eta}\;{\mu_0}^{2}.
\end{equation}

On comparing Eqs.(9),(18) and (19) we  see that a distribution of the dipole moments in an SQD ensemble significantly changes  the polarization Eq.(8) and Maxwell's wave equations  for an ensemble of quantum dots Eqs.(10) and (11) and, as a result, in a special case also the Beer's absorption length Eq.(18).

\end{document}